\documentclass[prb,twocolumn,preprintnumbers,superscriptaddress]{revtex4}

\usepackage{graphicx}
\usepackage{dcolumn}
\usepackage{amsmath}
\usepackage{color}

\begin{document}

\title{Electronic structure and magnetic correlations in trilayer nickelate superconductor La$_4$Ni$_3$O$_{10}$ under pressure}

\author{I. V. Leonov}
\affiliation{M. N. Mikheev Institute of Metal Physics, Russian Academy of Sciences, 620108 Yekaterinburg, Russia}
\affiliation{Institute of Physics and Technology, Ural Federal University, 620002 Yekaterinburg, Russia}

\begin{abstract}

It has been recently shown that under pressure trilayer Ruddlesden-Popper nickelate La$_4$Ni$_3$O$_{10}$ (LNO) becomes superconducting below a critical temperature $\sim$20~K, in addition to the infinite-layer and bilayer systems. Motivated by this observation, we explore the effects of electron correlations on its electronic structure and magnetic properties using the advanced DFT+dynamical mean-field theory approach. Our results for the normal-state electronic structure and correlation effects in LNO show much in common with the infinite-layer and bilayer nickelates, with remarkable site- and orbital-dependent renormalizations of the Ni $3d$ bands and notable incoherence of the Ni $d_{3z^2-r^2}$ states, caused by correlation effects. Our analysis of the Fermi surface and magnetic correlations suggests the emergence of competing spin and charge stripe states, implying the importance of in-plane spin fluctuations to explain superconductivity in this material. 

\end{abstract}

\maketitle

%%%%%%%%%%%%%%%%%%%%%%%
\section{Introduction}
%%%%%%%%%%%%%%%%%%%%%%%

%The theoretical and experimental description of the microscopic mechanisms of unconventional superconductivity is one of the most challenging problems in condensed matter physics. 
The recent discovery of superconductivity in the hole-doped ``infinite-layer'' nickelates $R$NiO$_2$ ($R$ -- rare earth element) has renewed great attention to the low-valence layered nickelates \cite{Li_2019,Hepting_2020,Zeng_2020,Osada_2021,Pan_2022}. 
In these mixed-valence systems, Ni ions adopt a square-planar NiO$_2$ structure, reminiscent of the electronic states of Cu$^{2+}$ ions in the hole-doped cuprate superconductors. Being isoelectronic to the parent cuprate superconductor CaCuO$_2$, with a critical temperature up to $T_c \sim 110$~K upon hole doping \cite{Azuma_1992}, the low-energy physics of hole-doped infinite-layer nickelates is dominated by the Ni $d_{x^2-y^2}$ electrons \cite{Lee_2004}. However, due to strong hybridization effects it is complicated by the presence of the rare-earth $5d$ states at the Fermi level, yielding a self-doped electronic structure and noncupratelike multiorbital Fermi surface (FS) \cite{Lee_2004,Kitatani_2020,Chen_2022a,Nomura_2022,Botana_2022}. 

Experimental and theoretical studies of the infinite-layer nickelates suggest a remarkable orbital-dependent localization of the Ni $3d$ states, implying the importance of strong correlations \cite{Kitatani_2020,Chen_2022a,Nomura_2022,Botana_2022}. Interestingly, superconductivity sets in at a mixed-valence state near to Ni$^{1.2+}$. While recent experiments provide evidence for sizable antiferromagnetic correlations, no static magnetic order has been detected yet \cite{Lin_2021,Zhou_2022,Lin_2022,Fowlie_2022}. Based on the resonant diffraction, the authors propose the formation of translational symmetry broken states (spin and charge density wave states) \cite{Rossi_2021,Tam_2021,Krieger_2021}, implying the complex interaction between spin and charge degrees of freedoms \cite{Si_2020,Werner_2020,Karp_2020a,Lechermann_2020b,Wang_2020,
Nomura_2020,Leonov_2020,Leonov_2021,
Lechermann_2022,Kreisel_2022,Pascut_2023,Kitatani_2023,
Slobodchikov_2022,Shen_2023,Chen_2022}.

In 2023, superconductivity with a high 
critical temperature $\sim$80~K has been experimentally determined in the bulk bilayer perovskite nickelate La$_3$Ni$_2$O$_7$, under pressure above 15 GPa \cite{Sun_2023,Hou_2023,Zhang_2023a,Liu_2023,Zhou_2023}. It belongs to the Ruddlesden-Popper (RP) series La$_{n+1}$Ni$_n$O$_{3n+1}$ with $n=2$. In contrast to the infinite-layer nickelates, in the bilayer system the Ni ions have a nominal Ni$^{2.5+}$ 
($3d^{7.5}$) electronic configuration. Moreover, superconductivity sets in near to a structural phase transition to the orthorhombic phase above $\sim$15 GPa, 
which is characterized by the absence of tilting of oxygen octahedra \cite{Sun_2023,Hou_2023,Zhang_2023a,Liu_2023}. 
Superconductivity is found to compete with a weakly insulating behavior at pressures below 15 GPa and a strange (bad) metal phase above $T_c$. Upon cooling below 153~K (at ambient pressure),
the bilayer nickelate shows anomaly in the transport properties associated with the formation of spin and charge 
density wave states \cite{Liu_2023}. Moreover, the experiments propose the significance of orbital-selective correlations in the bilayer system, consistent with the recent results of the 
electronic structure DFT+dynamical mean-field theory (DFT+DMFT) and GW+DMFT calculations \cite{Shilenko_2023,Fan_2023,Zhang_2023,Lechermann_2023,Christiansson_2023,
Yang_2023,Shen_2023b,Sakakibara_2023a,Cao_2023,Chen_2023,Wu_2023,Ryee_2023,Heier_2023}.

More recently, it has been demonstrated that under pressure above $\sim$15 GPa the trilayer RP ($n=3$) nickelate La$_4$Ni$_3$O$_{10}$ (LNO) reveals signatures of superconductivity with $T_c$ about 20-30 K \cite{Sakakibara_2023,Zhu_2023,Zhang_2023c,Li_2023,Li_2024}. It shares the same structural motif as that in the bilayer system (and high-$T_c$ cuprates) with the trilayer corner-sharing 
NiO$_6$ octahedra blocks separated by the La-O units. In contrast to the bilayer nickelate, it has (at least) two structurally distinct Ni sites: in the inner (Ni$_\mathrm{in}$) and in the outer (Ni$_\mathrm{out}$) NiO$_2$ layers of the trilayer blocks. Similarly to the bilayer system, under pressure above 15~GPa the trilayer nickelate LNO exhibits a structural phase transition to a high-symmetry 
tetragonal $I4/mmm$ phase, with no NiO$_6$ octahedra tilting. 

At ambient pressure, below $\sim$140~K LNO undergoes a metal-to-metal phase transition associated with the formation of unusual intertwined spin and charge density wave states \cite{Zhang_2020}. 
The pressure-driven transition is accompanied by the suppression of (long-range) density wave ordering, with the emergence of superconductivity above $\sim$15 GPa \cite{Sakakibara_2023,Zhu_2023,Zhang_2023c,Li_2023,Li_2024}. Moreover, 
experiments suggest the appearance of a strange metal behavior in the normal state of LNO extending up to room temperature \cite{Zhu_2023}. This implies the crucial importance of the effects of electronic correlations, Fermi surface nesting, and charge and spin density wave ordering to 
explain the properties of LNO. %{\color{red}Interestingly that the trilayer cuprates hold the highest $T_c$ among the high-$T_c$ cuprate superconductors, while in the trilayer nickelate $T_c$ is remarkably smaller than that in the bilayer LNO}. 
Nowadays, the effects of electronic correlations on the electronic structure, magnetism, and superconductivity in the trilayer nickelate are still poorly understood.

We address this issue in our present paper. In our work, we study the electronic structure and magnetic properties of the high-pressure $I4/mmm$ phase of the trilayer La$_4$Ni$_3$O$_{10}$ using the DFT+DMFT method \cite{Georges_1996,Kotliar_2006}. We explore the effects of electron correlations on the electronic structure and magnetic properties of this system. In our calculations, we adopt 
the experimental lattice parameters and atomic positions of LNO recently determined using synchrotron x-ray diffraction at about 40~GPa \cite{Li_2023}. 
In DFT we employ generalized gradient approximation with the Perdew-Burke-Ernzerhof (PBE) exchange functional as implemented in the Quantum ESPRESSO package \cite{Giannozzi_2017,DalCorso_2014}. 
Using DFT+DMFT we compute the normal-state electronic structure, Fermi surface, and magnetic correlations of the paramagnetic (PM) phase of LNO under pressure. In our calculations we employ the fully charge self-consistent DFT+DMFT method \cite{Pourovskii_2007} implemented with plane-wave pseudopotentials \cite{Leonov_2020b}. 
In order to take into account the effects of electronic correlations in the partially filled Ni $3d$ shell and charge transfer between the Ni $3d$, O $2p$, and La $5d$ 
valence states we construct a basis set of atomic-centered Wannier functions for the Ni $3d$, O $2p$, and La $5d$ orbitals using the energy window spanned by these bands \cite{Anisimov_2005}.

We use the continuous-time hybridization expansion quantum Monte-Carlo algorithm to treat the many-body and strong correlations effects in the Ni $3d$ shell \cite{Gull_2011}. The effects of electron correlations in the Ni $3d$ orbitals are treated by using the on-site Hubbard $U = 6$ eV and Hund’s exchange $J = 0.95$ eV, taken in accordance with previous studies of the infinite-layer and perovskite nickelates. The O $2p$ and La $5d$ valence states are uncorrelated and are treated on the DFT level within the self-consistent DFT+DMFT approach. We note that the tetragonal structure of LNO contains two structurally distinct Ni sites: one Ni$_\mathrm{in}$ and two equivalent Ni$_\mathrm{out}$ ions (one formula unit in the primitive cell). We therefore use a two-impurity-site extension of DFT+DMFT to treat the effect of correlations in the structurally distinct Ni sites. We employ the fully localized double-counting correction evaluated from the self-consistently determined local Ni $3d$ occupations. In order to compute the {\bf k}-resolved spectra and correlated Fermi surfaces we use Pad\'e approximants to perform analytic continuation of the self-energy results on the real energy axis. Using DFT+DMFT we study the normal-state electronic structure of PM LNO at a temperature $T = 290$ K (our calculations at $T=116$~K give quantitatively similar results).

%%%%%%%%%%%%%%%%%%%%%%%
\section{RESULTS AND DISCUSSION}
%%%%%%%%%%%%%%%%%%%%%%%

%%%%%%%%%%%%%%%%%%%%%
\subsection{Electronic structure}
%%%%%%%%%%%%%%%%%%%%%

\begin{figure}[tbp!]
\centerline{\includegraphics[width=0.5\textwidth,clip=true]{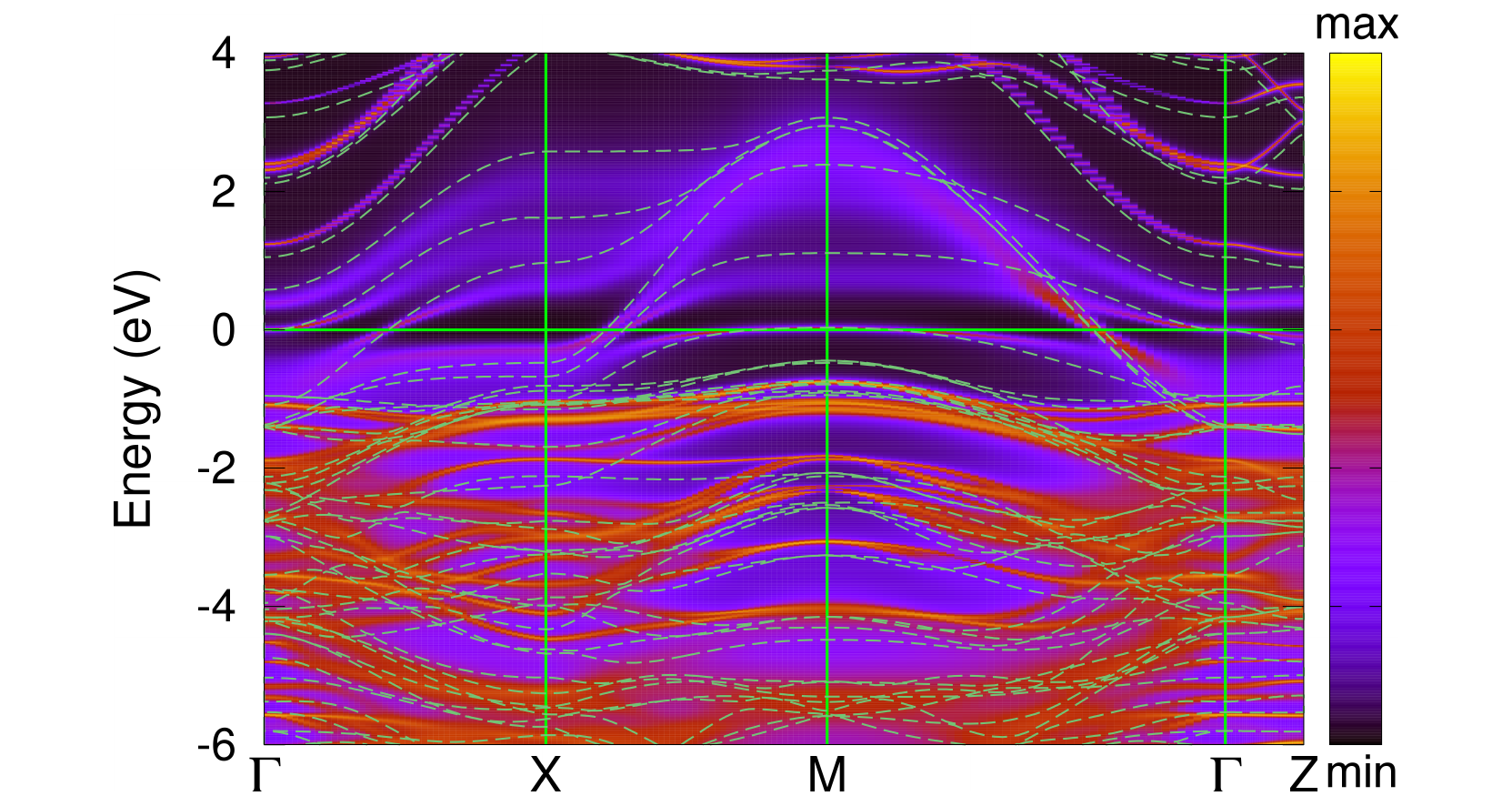}}
\caption{{\bf k}-resolved spectral functions of PM LNO obtained by DFT+DMFT at $T=290$ K. The nonmagnetic DFT (PBE) results are shown with green broken lines.}
\label{Fig_1}
\end{figure}

\begin{figure}[tbp!]
\centerline{\includegraphics[width=0.5\textwidth,clip=true]{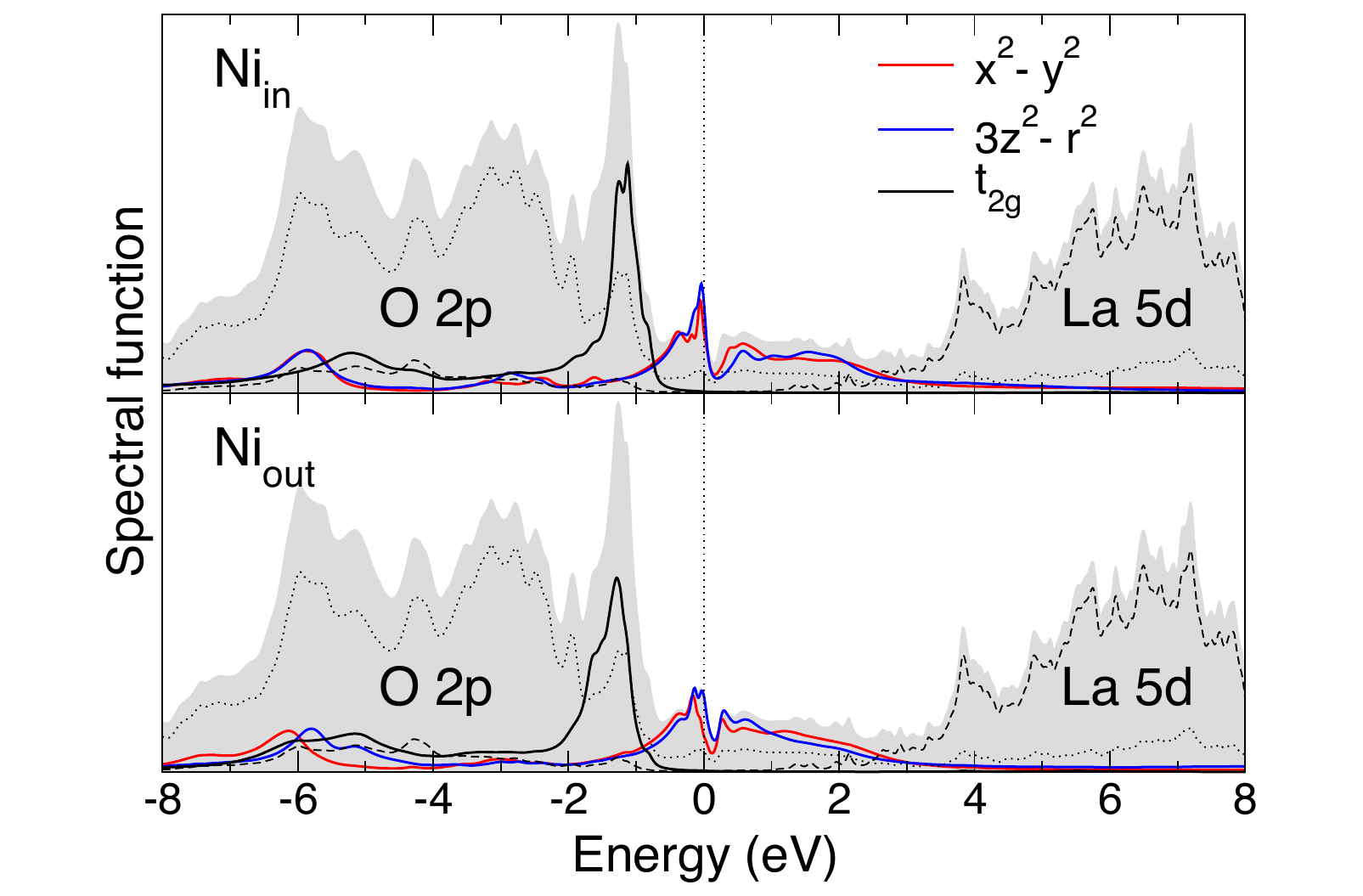}}
\caption{Orbital-dependent spectral functions of PM LNO obtained by DFT+DMFT at $T=290$ K. The partial Ni $t_{2g}$, $x^2-y^2$, and $3z^2-r^2$ orbital contributions are shown (per orbital). The partial Ni $3d$ states are magnified by a factor eight for better readability.}
\label{Fig_2}
\end{figure}

We begin with the electronic structure calculations of the high-pressure LNO. In Fig.~\ref{Fig_1} we display the ${\bf k}$-resolved spectral functions obtained using DFT+DMFT (for the crystal structure parameters taken at about 40 GPa). The corresponding orbital-dependent spectral functions are shown in Fig.~\ref{Fig_2}. Overall, the ${\bf k}$-resolved spectral functions of LNO remain similar to those obtained within the nonmagnetic DFT. Our results show a remarkable renormalization of the Ni $3d$ quasiparticle bands and incoherence of the spectral weight caused by correlation effects. The partially occupied Ni $d_{x^2-y^2}$ and $d_{3z^2-r^2}$ electronic states appear near the Fermi level and form a quasiparticle band between -2 eV and 3 eV. The Ni $t_{2g}$ states are fully occupied and are located at about -1.2 eV below the $E_F$. The occupied O $2p$ states appear at -8 to -0.6 eV below $E_F$, and exhibit strong hybridization with the Ni $3d$ and empty La $5d$ orbitals. The electronic states originating from the La $5d$ orbitals are empty and appear above 2 eV. {This picture qualitatively differs} from a self-doped electronic structure behavior of the infinite-layer systems \cite{Lee_2004}. 

We observe that the electronic states of the Ni ions in the inner and outer NiO$_2$ layers exhibit only minor differences. We note that the Ni$_\mathrm{in}$ $e_g$ orbitals and the Ni$_\mathrm{out}$ $d_{3z^2-r^2}$ orbitals show a quasiparticle peak at the Fermi level (at $T=290$ K), while the Ni$_\mathrm{out}$ $d_{x^2-y^2}$ states are close to form a pseudo gap at the $E_F$, with a quasiparticle peak located to about -0.14 eV below the Fermi level. Moreover, in close similarity to the bilayer system, the {\bf k}-resolved spectral functions of LNO show remarkably flat (and strongly incoherent) band dispersions originating from the Ni $d_{3z^2-r^2}$ orbitals located near the Fermi level at the Brillouin zone (BZ) M-point (note that in our previous work Ref.~\onlinecite{Shilenko_2023} in Fig.~1 the M and X points were mixed up by a mistake). In addition, the Ni $d_{x^2-y^2}$ orbitals form a nearly flat structure at about -0.14 eV below $E_F$ at the X point on the $\Gamma$-X branch, suggestive of a van Hove anomaly in the electronic spectra. Moreover, the same behavior is seen in the electronic structure of the infinite-layer and bilayer nickelates, while electronic correlations shift this anomaly closer to the $E_F$ (due to the Ni $3d$ bandwidth renormalization). Overall, the calculated electronic structure of LNO is very similar to that of the high-pressure bilayer nickelate, with two extra Ni $e_g$ quasiparticle bands to appear near the Fermi level \cite{Puggioni_2018,Li_2017,Jung_2022,Sakakibara_2023}. In particular, we observe a weakly dispersive band of the Ni $d_{3z^2-r^2}$ character to appear on the $\Gamma$-Z branch below the Fermi level, which crosses the $E_F$ near the $\Gamma$-point.

Similarly to that in the bilayer nickelate, the Wannier Ni $d_{x^2-y^2}$ and $d_{3z^2-r^2}$ orbital occupations in LNO are close to half filling, $\sim$0.53 and 0.56 (per spin-orbit) for 
the Ni$_\mathrm{in}$ and 0.55 and 0.57 for the Ni$_\mathrm{out}$ sites, respectively. That is, orbital polarization between the Ni $d_{x^2-y^2}$ and $d_{3z^2-r^2}$ orbitals is rather weak and 
bandwidth effects dominate over a potential Jahn-Teller orbital splitting (for Ni$^{3+}$ ions).
It is interesting to note that in the optimally doped $R$NiO$_2$ the Ni $d_{3z^2-r^2}$ orbital 
occupancy is considerably larger, $\sim$0.85 \cite{Leonov_2020,Leonov_2021}. In LNO the total Wannier Ni $3d$ occupations for the Ni$_\mathrm{in}$ and Ni$_\mathrm{out}$ ions are nearly the same, about 8.14 and 8.19, 
respectively. These values sufficiently differ from the formal oxidation state Ni$^{2.67+}$ ($3d^{7.33}$), implying the importance of negative charge transfer effects \cite{Zaanen_1985} (in contrast to the Mott-Hubbard regime of the infinite-layer systems \cite{Lee_2004,Kitatani_2020,Chen_2022a,Nomura_2022,Botana_2022}).
Our results for the instantaneous local magnetic moments $\sqrt{\langle \hat{m}^2_z \rangle}$ are 1.25 and 1.29 $\mu_\mathrm{B}$, respectively.
The fluctuating magnetic moments evaluated as  $M_\mathrm{loc} \equiv [k_\mathrm{B}T \int \chi(\tau) d\tau]^{1/2}$ (where $\chi(\tau) \equiv \langle \hat{m}_z(\tau)\hat{m}_z(0) \rangle $ is the local spin 
correlation function) are sufficiently smaller, about 0.36 and 0.43 $\mu_\mathrm{B}$.
Moreover, in accord with a mixed-valence Ni ion configuration our analysis of the weights of different 
atomic configurations of the Ni $3d$ electrons (being fluctuating between various atomic configurations within DMFT) gives 0.14, 0.54, and 0.29 
for the $d^7$, $d^8$, and $d^9$ configurations, respectively, for the Ni$_\mathrm{in}$ ions. For the Ni$_\mathrm{out}$ ions these are 0.11, 0.54, and 0.31, respectively. 
Our analysis of the spin-state configurations suggests strong interplay of the Ni $S=0$, $\frac{1}{2}$, and $1$ states in the electronic structure of LNO. Thus, 
the corresponding spin-state weights are 0.29, 0.41, and 0.28, in close similarity to that in the bilayer system.

%%%%%%%%%%%%%%%%%%%%%
\subsection{Orbital-selective behavior}
%%%%%%%%%%%%%%%%%%%%%

\begin{figure}[tbp!]
\centerline{\includegraphics[width=0.5\textwidth,clip=true]{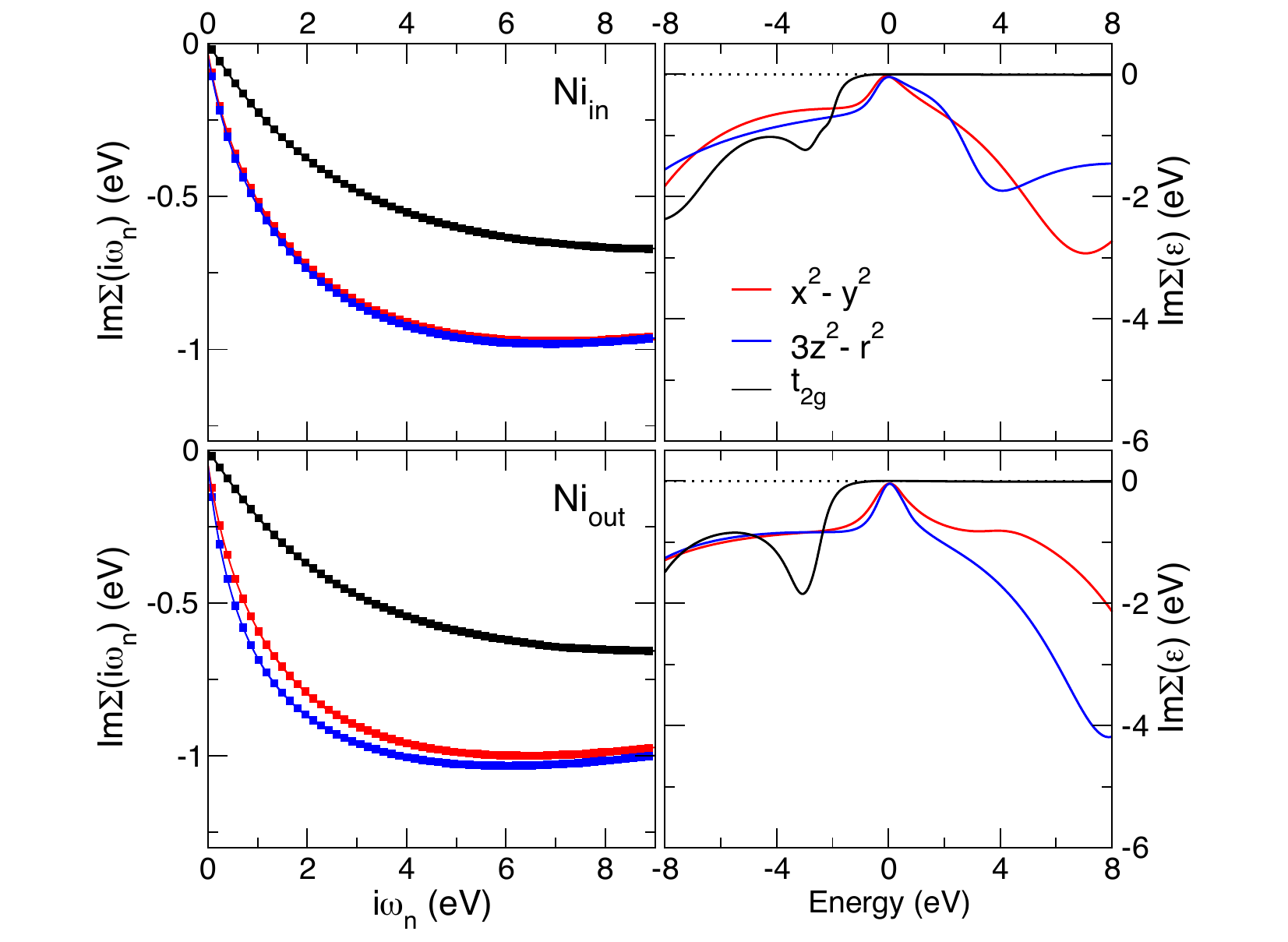}}
\caption{
Orbital-dependent Ni $3d$ self-energies $\mathrm{Im}[\Sigma(i\omega)]$ on the Matsubara axis and those analytically continued on the real energy axis $\mathrm{Im}[\Sigma(\omega)]$ for the crystallographically distinct Ni$_\mathrm{in}$ (top) and Ni$_\mathrm{out}$ (bottom) sites.}
\label{Fig_3}
\end{figure}

Next, we discuss the orbital-selective behavior and quasiparticle mass renormalizations of the Ni $3d$ bands in LNO under pressure. 
We observe that the electronic states of LNO obtained within DFT+DMFT show a remarkable renormalization of 
the quasiparticle Ni $3d$ bands and orbital-selective incoherence of the spectral weight (bad metal behavior). 
In particular, our analysis of the Ni $3d$ self-energies on the Matsubara contour suggests a typical Fermi liquid-like behavior 
with a substantial orbital-dependent quasiparticle damping of the Ni $3d$ states (see Fig.~\ref{Fig_3}). In fact, for the Ni$_\mathrm{in}$ ions $\mathrm{Im}[\Sigma(i\omega_n)]\sim 0.09$ and 0.11 eV 
for the Ni $d_{x^2-y^2}$ and $d_{3z^2-r^2}$ quasiparticle states at the first Matsubara frequency, at $T = 290$K. The fully occupied Ni $t_{2g}$ states are 
sufficiently coherent, with $\mathrm{Im}[\Sigma(i\omega_n)]$ below 0.02 eV at the first Matsubara frequency. At the same time, for the Ni$_\mathrm{out}$ ions 
the quasiparticle damping is noticeably higher, about 0.12 and 0.15 eV for the $d_{x^2-y^2}$ and $d_{3z^2-r^2}$ orbitals, respectively.
Using Pad\'e extrapolation of the self-energy $\Sigma(i\omega)$ to $i\omega \rightarrow 0$ we obtain 0.03 and 0.04 eV for Ni$_\mathrm{in}$, 
and $\sim$0.05 for the Ni$_\mathrm{out}$ $x^2-y^2$ and $3z^2-r^2$ states at the Fermi energy, respectively. 

In addition, we evaluate the quasiparticle 
mass enhancement factors {for the Ni $3d$ orbitals} as $m^*/m = [1 - \partial \mathrm{Im}[\Sigma(i\omega)]/ \partial i\omega]|_{i\omega\rightarrow 0}$.
Our DFT+DMFT calculations reveal significant correlation effects within the Ni $3d$ orbitals that suggest site- and orbital-dependent localization of the Ni $3d$ electrons. 
Our analysis of the quasiparticle mass enhancement for the Ni$_\mathrm{in}$ ions gives nearly orbital independent renormalization factors $m^*/m \sim 1.9$ for both the 
Ni $d_{x^2-y^2}$ and $d_{3z^2-r^2}$ orbitals. In contrast to this, for the Ni ions in the NiO$_2$ outer layer we obtain sufficient orbital differentiation, $\sim$2.1 and 2.5 
for the $d_{x^2-y^2}$ and $d_{3z^2-r^2}$ orbitals, respectively. That is, the $d_{3z^2-r^2}$ states of the Ni$_\mathrm{out}$ ions are seen to be more correlated and incoherent-like than 
the in-plane $d_{x^2-y^2}$ orbitals. The effective mass enhancement of the Ni $t_{2g}$ states is sufficiently weaker, $\sim$1.2 (for both Ni ions). 
This behavior implies the importance of site- and orbital-dependent localization of the Ni $3d$ electrons. Overall, the strength of correlation effects is seen to be sufficiently weaker than that in the bilayer {LNO. For the bilayer system with a nominal Ni$^{2.5+}$ valence state using DFT+DMFT with the same Hubbard $U=6$~eV and Hund's exchange $J=0.95$~eV we found $m^*/m \sim 2.3$ and 3 for the Ni $d_{x^2-y^2}$ and $d_{3z^2-r^2}$ orbitals, respectively.} 
It also differs qualitatively from the results for $R$NiO$_2$ which show a strong orbital differentiation of correlation effects with $m^*/m \sim 3$ 
and 1.3 for the Ni $d_{x^2-y^2}$ and $d_{3z^2-r^2}$ orbitals, respectively \cite{Leonov_2020,Leonov_2021}.

Overall, our estimates for the quasiparticle mass enhancement are in good agreement with experimental estimates ($\sim$2-2.6) \cite{Li_2017}. Our findings are also inline with a more narrow DFT bandwidth of the Ni $d_{3z^2-r^2}$ 
orbitals, which is by $\sim$17\% less than that for the $d_{x^2-y^2}$ states, for the Ni$_\mathrm{out}$ ions (although the Ni $x^2-y^2$/$3z^2-r^2$ orbital occupations are nearly the same, about 0.53/0.56 and 0.55/0.57 per spin orbit for the Ni$_\mathrm{in}$ and Ni$_\mathrm{out}$ ions, respectively). We also check how the correlation effects depend on the choice of the 
Hubbard $U$ and Hund's exchange $J$. In fact, it was found that the effective mass renormalization of the Ni $3d$ states depends sensitively upon variation of 
the $U$ and $J$ values. 
%Thus, for $U=4$ eV and $J=0.45$ eV we obtain $m^*/m \sim 1.4$ and 1.5 for the Ni $x^2-y^2$ and $3z^2-r^2$ orbitals for the Ni$_\mathrm{out}$ ions, respectively. For $U=6$ eV and $J=0.45$ eV we obtain $\sim$1.6 and 1.8, for $U=4$ eV and $J=0.95$ eV $\sim$1.6 and 1.9, and for $U=8$ eV and $J=0.95$ eV $\sim$2.7 and 3.1, respectively. 
Our results for the Ni $x^2-y^2$/$3z^2-r^2$ orbital occupations and $m^*/m$ obtained for different Hubbard $U$ and Hund's exchange $J$ values are summarized in Table~\ref{tab_1}.
In particular, our estimates 
reveal a remarkable sensitivity of $m^*/m$ upon variation of the Hund's exchange $J$, while dependence on the $U$ value is much weaker. This behavior is consistent with that 
obtained for the bilayer system. This implies the importance of multi-orbital Hund's metal physics to explain the electronic properties of LNO. At the same time, variations of the interaction parameters $U$ and $J$ do not affect much the Ni $x^2-y^2$ and $3z^2-r^2$ orbital occupations.

Moreover, our DFT+DMFT calculations for the electron-doped trilayer La$_4$Ni$_3$O$_{10}$ (under pressure) with a nominal valence state Ni$^{2.5+}$ as that in the bilayer 327 LNO (to mimic the filling in La$_3$Ni$_2$O$_7$) reveal a notable enhancement and orbital-selectivity of quasiparticle mass renormalizations $m^*/m$, which become close to that of the bilayer LNO (see Table~\ref{tab_2}). In agreement with this, for the hole-doped bilayer LNO with a nominal Ni$^{2.67+}$ state, we note a sizable decrease of correlation effects, with $m^*/m \sim 2.1$ and 2.8 for the Ni $x^2-y^2$ and $3z^2-r^2$ orbitals, respectively. (Here to treat the effects of doping on the electronic structure of the tri- and bilayer nickelates we employ a rigid-band approximation within DFT+DMFT).

\begin{table}[h]
\caption{Site-dependent Ni $d_{x^2-y^2}$ and $d_{3z^2-r^2}$ Wannier orbital occupations (per spin-orbit) and quasiparticle mass renormalizations $m^*/m$ of the trilayer LNO (with a crystal structure taken under pressure of $\sim$40 GPa) calculated by DFT+DMFT for different Hubbard $U$ and Hund's exchange $J$ values (in eV). }
\begin{ruledtabular}
\begin{tabular}{lcccccc}
\multicolumn{1}{l}{Ni ion} & \multicolumn{1}{c}{$U$} & \multicolumn{1}{c}{$J$} & \multicolumn{2}{c}{orbital occupations} & \multicolumn{2}{c}{$m^*/m$} \\
& & & $x^2-y^2$ & $3z^2-r^2$ & $x^2-y^2$ & $3z^2-r^2$ \\
\hline
& 4 & 0.45 & 0.53 & 0.57 & 1.3 & 1.4    \\
& 4 & 0.95 & 0.53 & 0.56 & 1.6 & 1.6   \\
Ni$_\mathrm{in}$ & 6 & 0.45 & 0.52 & 0.57 & 1.6 & 1.6    \\
& 6 & 0.95 & 0.53 & 0.56 & 1.9 & 1.9    \\
& 8 & 0.95 & 0.53 & 0.55 & 2.4 & 2.4   \\
\hline
& 4 & 0.45 & 0.56 & 0.58 & 1.4 & 1.5    \\
& 4 & 0.95 & 0.56 & 0.58 & 1.6 & 1.9    \\
Ni$_\mathrm{out}$ & 6 & 0.45 & 0.55 & 0.58 & 1.6 & 1.8   \\
& 6 & 0.95 & 0.55 & 0.57 & 2.1 & 2.5     \\
& 8 & 0.95 & 0.55 & 0.57 & 2.7 & 3.1   \\
\end{tabular}
\end{ruledtabular}
\label{tab_1}
\end{table}

\begin{table}[h]
\caption{Ni $d_{x^2-y^2}$ and $d_{3z^2-r^2}$ Wannier orbital occupations (per spin-orbit) and quasiparticle mass renormalizations $m^*/m$ for the electron-doped La$_4$Ni$_3$O$_{10}$ (under pressure) with a nominal valence state Ni$^{2.5+}$ (as that in the pure bilayer 327 LNO) (Ni$_\mathrm{in}$/Ni$_\mathrm{out}$). Our results for the hole-doped bilayer La$_3$Ni$_2$O$_7$ (under pressure) with a nominal valence state Ni$^{2.67+}$ (as that in the pure trilayer LNO) are depicted as Ni$^{2.67+}$. The DFT+DMFT calculations are performed for the Hubbard $U=6$ eV and Hund's exchange $J=0.95$ eV, at a temperature $T=290$ K. To treat the effects of doping on the electronic structure of the tri- and bilayer nickelates we employ a rigid-band approximation within DFT+DMFT.}
\begin{ruledtabular}
\begin{tabular}{lccccc}
\multicolumn{1}{l}{Ni ion} & \multicolumn{1}{c}{valence} &  \multicolumn{2}{c}{orbital occupations} & \multicolumn{2}{c}{$m^*/m$} \\
& & $x^2-y^2$ & $3z^2-r^2$ & $x^2-y^2$ & $3z^2-r^2$ \\
\hline
Ni$_\mathrm{in}$ &  2.5+ &  0.54 & 0.57 & 2.1 & 2.8    \\
Ni$_\mathrm{out}$ & 2.5+ &  0.57 & 0.59 & 2.3 & 2.7    \\
\hline
Ni & 2.67+ & 0.53 & 0.59 & 2.1 & 2.8    \\
\end{tabular}
\end{ruledtabular}
\label{tab_2}
\end{table}

Our results agree well with our analysis of the orbital-dependent local spin susceptibility $\chi(\tau)$, 
evaluated within DFT+DMFT (see Figs.~S1 and S2 in Supplemental Material~\cite{suppl}). Our calculations suggest the proximity of the Ni $e_g$ states to orbital-selective localization. Moreover, the Ni $3d$ electron localization has a pronounced site selectivity. In fact, the Ni$_\mathrm{out}$ $e_g$ 
states show more localized behavior than those for the Ni$_\mathrm{in}$ ions. In particular, for the Ni$_\mathrm{out}$ ions the $d_{3z^2-r^2}$ orbitals show a slow decaying 
behavior of $\chi(\tau)$ to 0.03 $\mu_\mathrm{B}^2$ at $\tau=\beta/2$, while for the $d_{x^2-y^2}$ states it is $\sim$0.02 $\mu_\mathrm{B}^2$. For the Ni$_\mathrm{in}$ 
ions this orbital selective behavior is less pronounced, $\chi(\tau)$ of about 0.01 $\mu_\mathrm{B}^2$ at $\tau=\beta/2$ for both the $d_{3z^2-r^2}$ and $d_{x^2-y^2}$ orbitals.
Our results therefore suggest that magnetic correlations in LNO are at the verge of a site- and orbital-dependent formation of local magnetic moments, suggestive of Hund's metal behavior \cite{Georges_2013}.

%%%%%%%%%%%%%%%%%%%%%
\subsection{Fermi surface and nesting}
%%%%%%%%%%%%%%%%%%%%%

\begin{figure}[tbp!]
\centerline{\includegraphics[width=0.5\textwidth,clip=true]{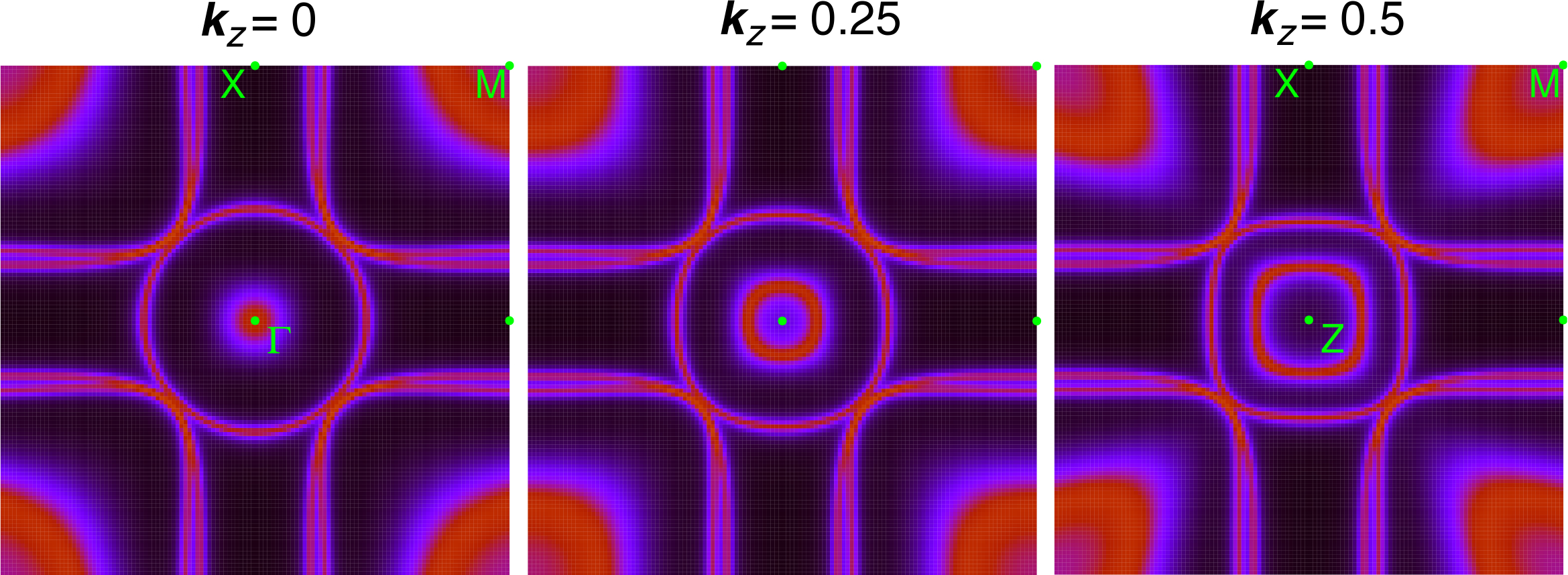}}
\caption{
Our results for the correlated Fermi surfaces for different $k_z$ calculated by DFT+DMFT at $T=290$ K.}
\label{Fig_4}
\end{figure}

Next, we compute the correlated FS using DFT+DMFT at $T=290$ K. To this end, we evaluate the ${\bf k}$-dependent spectral 
weight $A({\bf k},\omega)=-1/\pi~\mathrm{Im}[\omega+\mu-H_{\bf k}-\Sigma(\omega)]^{-1}$ at $\omega=0$. Here, $H_{\bf k}$ is the self-consistent Kohn-Sham Hamiltonian in the Wannier basis set, and $\Sigma(\omega)$ is
the self-energy analytically continued on the real energy axis. In Fig.~\ref{Fig_4} we show our results for the in-plane quasiparticle FSs for different $k_z$. 
Our results for the electronic structure and FSs obtained by DFT+DMFT for different $U$ and $J$ are shown in SM~\cite{suppl}.
In agreement with previous angle-resolved
photoemission spectroscopy measurements, we obtain a large hole FS that closely resembles the FS of optimally 
hole-doped cuprates, including its $x^2-y^2$ orbital character, hole occupation level, and strength of correlation effects \cite{Li_2017,Jung_2022}. 
We note that correlation effects mainly result in the orbital-dependent quasiparticle mass renormalizations and incoherence of the spectral weight, while the FS topology 
is notably similar to that obtained within DFT (see Fig.~S7).
Moreover, our results reveal a remarkable flattening of the electronic band structure and shift of the spectral weight below the $E_F$ towards the Fermi level at the $\Gamma$-X and $\Gamma$-Z branches of the BZ, driven by correlation effects (see SM \cite{suppl}).

Our results for the FSs bear a striking resemblance to those of the bilayer nickelate (see, e.g., Refs.~\onlinecite{Shilenko_2023,Lechermann_2023}).
We observe two electron FS pockets centered at the BZ $\Gamma$ (Z) point and three (two of which are nearly degenerate) hole pockets at the M point. 
Overall, the quasiparticle FSs are two-dimensional, except for the internal FS sheet predominantly of the Ni$_\mathrm{out}$ $d_{3z^2-r^2}$ character centered at the $\Gamma$ point, along the $k_z$ axis. 
The latter shows a remarkable dependence on $k_z$, forming a hourglass-like FS. It is associated with a weakly dispersive Ni$_\mathrm{out}$ $d_{3z^2-r^2}$ 
band on the $\Gamma$-Z branch, crossing the $E_F$ near the $\Gamma$ point (see Fig.~\ref{Fig_1}). 

We note that the FSs exhibit strong orbital-dependent incoherence caused by correlation effects. 
It is particularly notable for the hole pocket near the M point originating from the shallow flat band of the Ni $d_{3z^2-r^2}$ character located near $E_F$.
This FS pocket shows a remarkable $k_z$ dispersion (a large change of the size with $z$). We also notice a variation of the shape of the 
large electron FS pocket of the Ni $d_{x^2-y^2}$ orbital character near the $\Gamma$ (Z) point, from a circle-like at $k_z=0$ to a square-like at $k_z=1/2$ ($\pi/z$), 
with no change of its size. In comparison to the bilayer system, our results indicate a notably reduced two-dimensionality of the electronic structure. 

Moreover, our analysis of the FS suggests its high sensitivity towards hole or electron doping, resulting in a reconstruction of the FS topology (a Lifshitz transition). Thus, hole doping tends to suppress the internal FS sheet centered at the BZ $\Gamma$ (Z) point. In contrast, by doping with electrons one would diminish the strongly incoherent hole FS sheet near the M point. It affects both FS dimensionality and FS nesting. All this suggests high sensitivity of the properties of LNO to stoichiometry and doping.

%%%%%%%%%%%%%%%%%%%%%
\subsection{Magnetic correlations}
%%%%%%%%%%%%%%%%%%%%%

\begin{figure}[tbp!]
\centerline{\includegraphics[width=0.5\textwidth,clip=true]{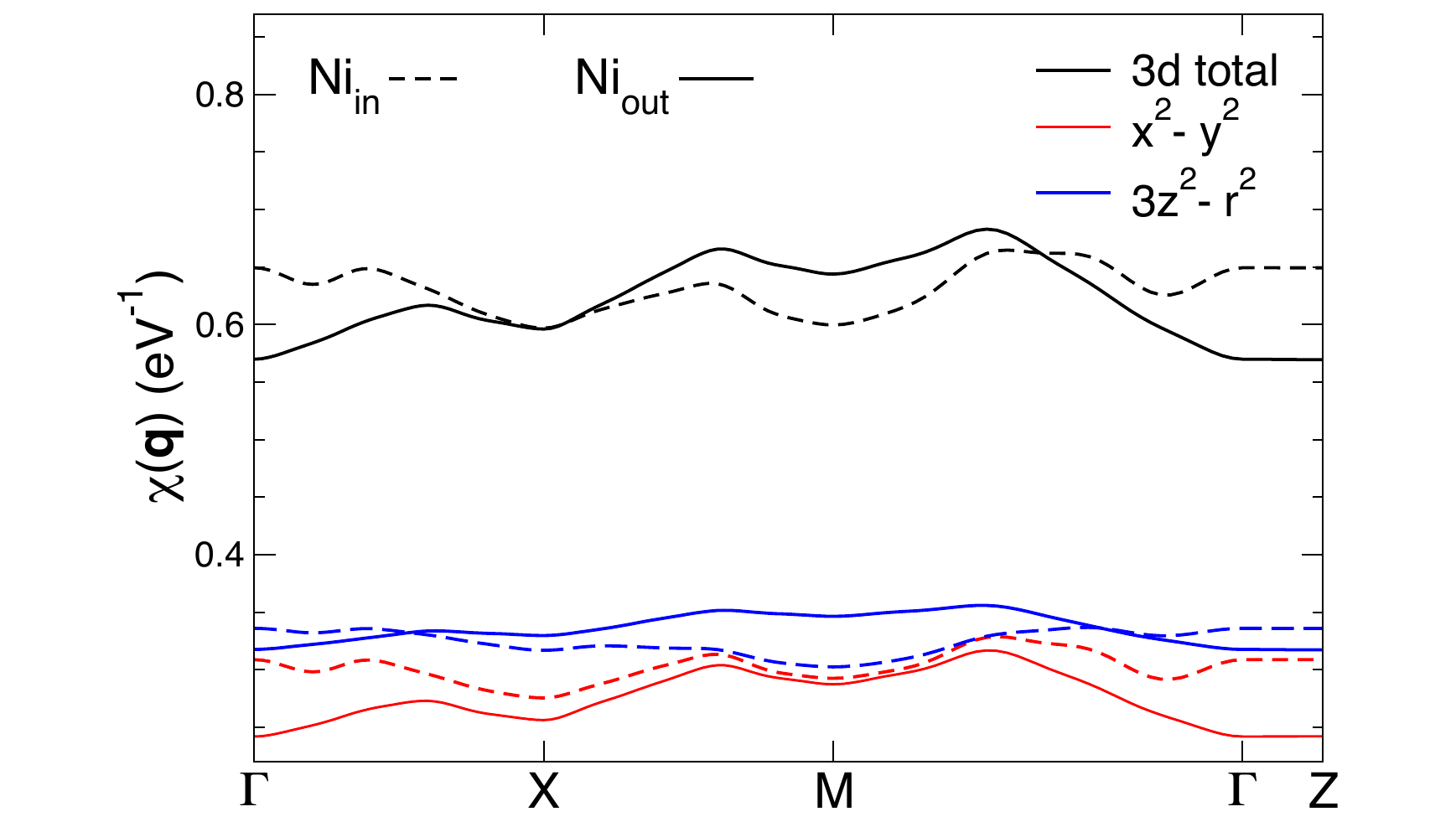}}
\caption{
Orbitally resolved static spin susceptibility $\chi({\bf q})$ of LNO obtained by DFT+DMFT at $T=290$ K.}
\label{Fig_5}
\end{figure}

Obviously, the correlated Fermi surfaces of the high-pressure LNO exhibit multiple in-plane nesting effects (see Fig.~\ref{Fig_4}), 
implying the presence of competing ordering effects in LNO.
Therefore, we compute the momentum-dependent static magnetic susceptibility $\chi({\bf q})$ in the particle-hole bubble approximation within DFT+DMFT as 
$\chi({\bf q})=-k_BT\mathrm{Tr} \Sigma_{{\bf k}, i\omega_n}G_{\bf k}(i\omega_n)G_{{\bf k}+{\bf q}}(i\omega_n)$. $G_{\bf k}(i\omega_n)$ is the local \emph{interacting} Green's function for the Ni $3d$ states evaluated on the Matsubara contour $i\omega_n$. In Fig.~\ref{Fig_5} we display our results for the site- and orbital-dependent contributions to $\chi({\bf q})$. 
Our results for $\chi({\bf q})$ obtained by DFT+DMFT for different parameters $U$ and $J$ are given in SM \cite{suppl}.
$\chi({\bf q})$ exhibits complex structure with multiple well-defined maxima, suggesting a complex interplay between spin and charge density wave states in LNO. Our results for the Ni$_\mathrm{out}$ ions exhibit three well-defined maxima at an incommensurate wave vector $(0.3~0~0)$ on the $\Gamma$-X, $(0.5~0.31~0)$ on the X-M, and $(0.31~0.31~0)$ on the $\Gamma$-M branches of the BZ. The most pronounced instability is associated with that on the $\Gamma$-M branch, competing with two minor instabilities at $(0.5~0.31~0)$ and $(0.3~0~0)$. Most importantly, this result agrees qualitatively well with that obtained for the bilayer system. In La$_3$Ni$_2$O$_7$ $\chi({\bf q})$ also shows a similar structure. This is due to the fact that the FS topology (and hence nesting effects) in LNO overall resembles that in the bilayer nickelate. 
%Moreover, our analysis of the previously published results for the optimally doped infinite-layer nickelates suggests qualitatively similar structure of $\chi({\bf q})$.
All this supports the picture of multiple (or intertwined) spin and change density wave instability driven by the FS in these nickelates. The later can in principle compete with the super-exchange interactions, giving the complex interplay between electron correlations, spin and charge stripe states. 

While the electronic states of the Ni sites in the inner and outer NiO$_2$ layers show minor quantitative differences, $\chi({\bf q})$ reveals sufficiently distinct behavior for the Ni$_\mathrm{in}$ ions with different magnetic correlations. For the Ni$_\mathrm{in}$ ions $\chi({\bf q})$ shows a rather complex behavior with multiple structures, distinct from those in the double layer system. For Ni$_\mathrm{in}$ ions the most pronounced instability is associated with a vector close to $(0.29~0.29~0)$. Particularly interesting is the importance of a ferromagnetic instability (at the $\Gamma$ point).
It is worth noting a rather sensitive dependence of $\chi({\bf q})$ upon a variation of the Hubbard $U$ and Hund's exchange $J$ values. Thus, for the relatively small for Ni ion $U=4$ eV and $J=0.45$ eV $\chi({\bf q})$ is found to exhibit a major anomaly at $\mathbf{q}=\Gamma$ for the Ni$_\mathrm{in}$ ions, resulting in the appearance of a ferromagnetic instability. On the other hand, correlation effects are found to suppress this anomaly with a pronounced spin (or charge) density wave instability for the Ni$_\mathrm{out}$ sites, with a main anomaly at the $\Gamma$-M and $\Gamma$-X branches of the BZ (see SM~\cite{suppl}).

Our results suggest the complex interplay between spin and charge states in the high pressure phase of LNO, which can be important for understanding of superconductivity in low-valence nickelates. We note that at ambient pressure LNO exhibits the formation of unusual spin and charge density wave states below a metal-to-metal phase transition. Moreover, the infinite-layer systems \cite{Rossi_2021,Tam_2021,Krieger_2021}, hole-doped mixed-valence nickelates (La,Sr)$_2$NiO$_4$ (with Sr $x = 1/3$, Ni$^{2.33+}$ ions) and square-planar La$_4$Ni$_3$O$_8$ with Ni$^{1.33+}$ and La$_3$Ni$_2$O$_6$ with Ni$^{1.5+}$ ions all reveal a similar behavior \cite{Lee_1997, Bernal_2019,Zhang_2019,Zhang_2020}. For hole-doped $R$NiO$_2$ the DFT+DMFT calculations suggest the possible formation of charge-disproportionated (bond-disproportionated) striped states \cite{Slobodchikov_2022,Shen_2023,Chen_2022}. All this suggests the importance of density wave states (and most importantly their fluctuations) to explain the electronic properties of LNO \cite{Tranquada_1995,Keimer_2015} . Superconductivity observed in LNO can be associated with the suppression of charge and spin density wave ordering under pressure, which leads to a sharp increase of spin fluctuations. This implies the importance of in-plane spin fluctuations to explain superconductivity in LNO under pressure. 

%%%%%%%%%%%%%%%%%%%%%
\section{Conclusion}
%%%%%%%%%%%%%%%%%%%%%

In conclusion, using the DFT+DMFT approach we studied the normal-state electronic properties of the trilayer nickelate superconductor La$_4$Ni$_3$O$_{10}$. We note that the electronic structure and correlation effects in the trilayer system share the common features of the infinite-layer and bilayer nickelates. In agreement with experimental estimates, our results show a remarkable renormalization of the Ni $3d$ bands, implying the site- and orbital-selective localization of the Ni $3d$ electrons. Our analysis suggests a high sensitivity of the properties of LNO to stoichiometry and doping. Our results imply the emergence of competing stripe states on a microscopic level, which affect the electronic structure and superconductivity of this material. We propose the presence of strong in-plane spin fluctuations, associated with the suppression of charge and spin density wave ordering in the trilayer LNO under pressure.

%%%%%%%%%%%%%%%%%%%%%
%\section{ACKNOWLEDGMENTS}
%%%%%%%%%%%%%%%%%%%%%

\begin{acknowledgments}
We acknowledge the support of the Ministry of Science and Higher Education of the Russian Federation, project No. 122021000038-7 (theme ``Quantum'' ).

\end{acknowledgments}

\end{document}